\documentclass[10pt,twocolumn]{article} 
\usepackage{latex8}
\usepackage{times}
\usepackage{graphicx}
\usepackage{subfigure}
\usepackage{amsfonts}
\usepackage{epsfig}
\usepackage{multirow}

\begin{document}

\author{
Damiano Bolzoni, Sandro Etalle, Pieter Hartel\\
\textit{University of Twente,}\\
\textit{Distributed and Embedded System Group,}\\
\textit{P.O. Box 2100, 7500 AE Enschede, The Netherlands}\\
\textit{\{damiano.bolzoni, sandro.etalle, pieter.hartel\}@utwente.nl}\\
\and
Emmanuele Zambon\\
\textit{Universita' Ca'Foscari di Venezia,}\\
\textit{Dipartimento di Informatica,}\\
\textit{Via Torino 155, 30172 Mestre (VE), Italy}\\
\textit{ezambon@dsi.unive.it}\\
}

\title{
POSEIDON: a 2-tier Anomaly-based Network Intrusion
      Detection System
      \thanks{This research is supported by the
    research program Sentinels (http://www.sentinels.nl). Sentinels is
    being financed by Technology Foundation STW, the Netherlands
    Organization for Scientific Research (NWO), and the Dutch Ministry
    of Economic Affairs.}}

\maketitle

\thispagestyle{empty}

\begin{abstract}
  We present POSEIDON, a new anomaly-based network intrusion detection system.
  POSEIDON is payload-based, and has a two-tier architecture: the
  first stage consists of a Self-Organizing Map, while the second one
  is a modified PAYL system. Our benchmarks on the 1999 DARPA data set show a higher detection 	  rate and lower number of false positives than PAYL and PHAD.
\end{abstract}

\section{Introduction}
Intrusion detection systems were introduced by Anderson ~\cite{And80}
and formalized later by Denning \cite{Den87}. Nowadays, there exist
two main types of network intrusion detection methods:
\emph{anomaly-based} and \emph{signature-based}. In signature-based
methods, (e.g.~Snort \cite{Roe99,snort}) a characteristic trait of the
intrusion is developed off-line, and then loaded in the intrusion
database before the system can begin to detect this particular
intrusion. This usually yields good results in terms of low false
positives, but has drawbacks: firstly in most systems, \emph{all} new attacks will
go unnoticed until the system is updated, creating a window of
opportunity for attackers to gain control of the system under attack.
Secondly, only known attacks can be detected, and while this could
be acceptable for detecting attacks to e.g., the OS, it makes it much
harder to use signature-based system for protecting web-based services,
because of their ad-hoc nature. Notably, the protection of
web-services is becoming a high-impact problem~\cite{isc}.

Anomaly-based systems (ABS), on the other hand, build statistical
models that describe the normal behaviour of the network, and flag any
behaviour that significantly deviates from the norm as an attack.
This has the advantage that new attacks will be detected as soon as
they take place. ABS can be applied also to ad-hoc networked systems such as web-based services. The
disadvantage is that ABS needs an extensive model building phase: a
significant amount of data (and thus a significant period of time) is
needed to build accurate models of legal behaviour.

Most network intrusion detections systems in use today are signature-based,
however, new attacks are devised with increasing frequency every day
(see~\cite{isc} for weekly and monthly single attack rates), so anomaly-based
systems become increasingly attractive.

Every network intrusion detection system suffers from (1) false positives
(false alarms), in which legal behaviour is incorrectly flagged as an
attack and (2) false negatives, or misses, in which true attacks are
undetected. Anomaly-based systems are more vulnerable to these
problems than signature-based systems because they use statistical
models to detect intrusions.

ABS can extract information to detect attacks from different layers:
packet headers, packet payload or both. \emph{Header information} is
mainly useful to recognize attacks aiming at vulnerabilities of the network stack implementation or probing the operating system to identify
active network services. On the other hand, \emph{payload information}
is most useful to identify attacks against vulnerable applications
(since the connection that carries the attack is established in a
normal way)~\cite{WS04}. Without pretending to be globally better than
other types of ABS, payload-based systems have importance of their
own, as they are particularly suitable for detecting popular attacks
such as those on the HTTP protocol, and worms (see Wang and
Stolfo~\cite{WCS05} and Costa et al.~\cite{CCCR+05} for a discussion). Notably, PAYL and the system of
Kruegel et al.~\cite{KTK02} are mainly payload-based, while PHAD
\cite{MC02} is partly payload based.

\begin{figure*}[ht]
\centering
\subfigure[PAYL]
{
        \includegraphics[width=15cm]{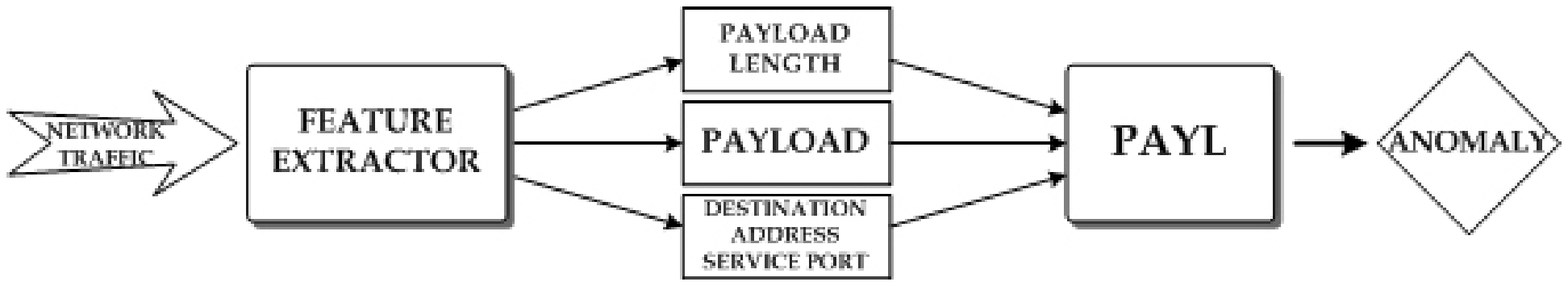}
}
\subfigure[POSEIDON]
{
        \includegraphics[width=15cm]{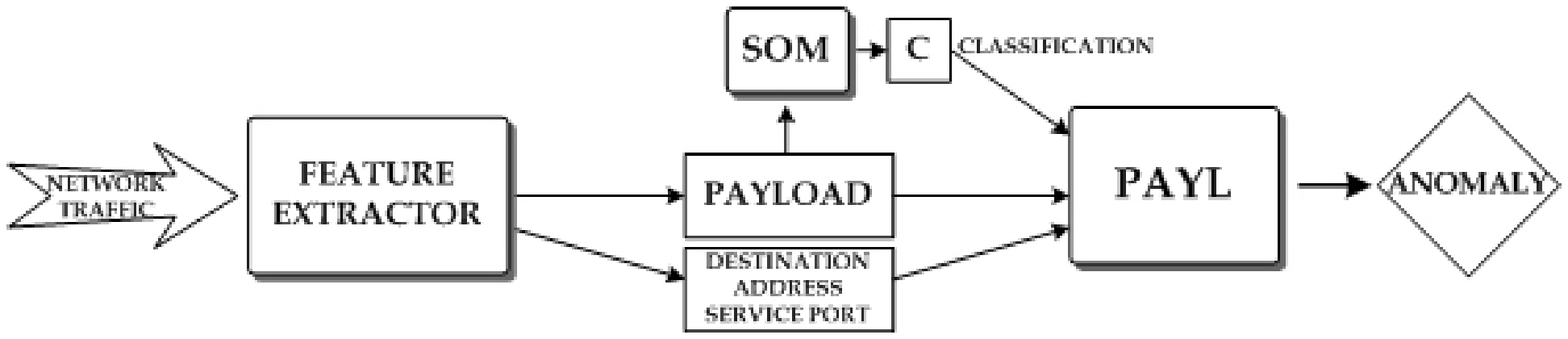}
}
\caption{PAYL and POSEIDON architectures\label{fig:architecture}}
\end{figure*}

\paragraph{Contribution} 
In this paper we propose POSEIDON (Payl Over Som for Intrusion
DetectiON): a two-tier network intrusion detection architecture. The first tier consists of a self-organizing map (SOM), and is used exclusively
to classify payload data; the second tier consists of a slight
modification of the well-known PAYL system~\cite{WS04} (see Figure
\ref{fig:architecture}).

POSEIDON is payload-based: it uses only destination address and
service port numbers to build a profile for each port monitored, and
it does \emph{not} consider other header features.

We have extensively benchmarked our system w.r.t.~PAYL \cite{WS04} (also
by replicating the PAYL experiments) and PHAD \cite{MC02} using the 1999
DARPA benchmark~\cite{LHFK+00}. PAYL and PHAD are the reference ABS
based on payload analysis. On this data set, our experiments show:

\begin{itemize}
\item a \emph{higher} detection rate and \emph{lower} number of
  false positives than PAYL and PHAD.
\item a reduction of the number of profiles used w.r.t.~PAYL. This has a
  positive influence on the runtime efficiency of the system.
\end{itemize}

Incidentally, being payload-based, our system takes into consideration
only what Mahoney and Chan \cite{MC03} call the \emph{legitimate} data
of the 1999 DARPA data set, implying that we can legitimately expect
that the system in real life performs as well as it does on the DARPA
benchmark.

Let us now explain the reasons that brought us to the development of this architecture. First of all, for the classification phase, we believe that a self-organizing map - in general - can yield to a high quality classification, i.e.~clusters with a high intra-cluster similarity and high inter-cluster dissimilarity, without having to take into account the length of the packet. This can be used to build good profiles.

At the same time, we believe that a SOM is not as effective when it comes to the detection phase, i.e.~to finding whether a given packet is anomalous w.r.t.~the cluster it has been classified in. In a SOM, the detection phase is accomplished by comparing the current packet quantization error with matching cluster quantization error: this method can be heavily influenced by payload byte order, because it is based on a distance function. For the detection, we believe that the n-gram algorithm used by PAYL is more suitable. 

On the other hand, we believe that the Achilles' heel of the PAYL architecture lies in the classification it adopts: the algorithm uses packet payload length information to classify packets and thus to define clusters. This, together with the fact that - for efficiency reasons - clusters have to be merged, yields in our opinion to a too low \textit{intra-cluster} similarity: two packets belonging to the same cluster can present very different byte distribution, without that this indicates an attack. 

By combining a SOM with the n-gram algorithm we obtained an architecture that combines the advantages of the SOM (the realization of clusters with high intra-cluster similarity) with those of PAYL (the ability to detect when a packet is anomalous w.r.t.~a given cluster). 
The results we have obtained on the DARPA substantiate our beliefs.

This paper is structured as follows: Section~\ref{sec:architecture}
presents the internals of POSEIDON and of PAYL; in
Section~\ref{sec:results} we describe benchmarking experiments and
compare obtained results with PAYL and PHAD. In
Section~\ref{sec:related} we discuss other related work. Finally, in
Section~\ref{sec:conclusion} we draw our conclusions and set the
course for further developments. In the appendix we report the
pseudo-code of POSEIDON.

\section{Architecture\label{sec:architecture}}

Network intrusion detection systems are either \emph{packet-oriented} or
\emph{connection-oriented}. In the former architecture, every packet
is analysed as soon as it arrives, without trying to correlate it with
previous collected data. On the other hand, connection-oriented
systems work either by (a) reassembling the whole connection (commonly
only from client to server) - waiting until the connection is closed -
to analyse the connection payload, or (b) by gathering statistics which
consider, e.g., the amount of bytes transmitted and received, the
duration of the connection, the protocol type and final connection
status.

POSEIDON, like most network intrusion detection systems, is packet-oriented.
This architecture presents two main advantages: firstly, POSEIDON can
identify and block an attack \emph{while} it is taking place
(intrusion prevention). Secondly, connection-based
systems are computationally more expensive, in particular they require
a huge amount of memory resources to keep all the segments to analyse.
This makes connection-based system more suitable for off-line
analysis. On the other hand, connection-based systems support a
finer-grained analysis.

Our starting point is the PAYL architecture. Our algorithm receives as
input a packet and \emph{classifies} the packet, without prejudice for
any of its properties, such as length, destination port or application
data semantics. The idea is that the classifier keeps as much
information as possible about packets (e.g.\ high-dimensional data)
for the anomaly detection phase: we also want the classifier to
operate in an unsupervised manner. This is a typical clustering
problem which can be properly tackled using neural networks in general
and Self-Organizing Maps (SOM)~\cite{Koh95} in particular. SOMs have been widely used in the past both to classify network data
and to find anomalies. Here, we use them for pre-processing.

Our architecture combines a SOM with a modified PAYL
algorithm.  Figure \ref{fig:architecture} shows a comparison between
our architecture and PAYL's.

We now give a high-level description of the algorithms underlying our
system, a more formal description is reported in the appendix. We
first describe the SOM. Later in the section, we introduce PAYL,
focusing on the main differences between our approach and the PAYL
approach towards classification of network data.

\subsection{SOM classification model\label{sec:SOM}}

Self organizing maps are defined as topology-preserving single-layer
maps in which the topological structure, imposed on the nodes in the
network, is not changed during classification (preserving
neighbourhood relations) and there is only one layer of nodes. A SOM
is suitable to analyse high-dimensional data and belongs to the
category of competitive learning networks~\cite{Koh95}. Nodes are also
called \emph{neurons}, to remind us of the artificial intelligence
nature of the algorithm. Each neuron $n$ has a \emph{vector of
  weights} $w_n$ associated to it: the dimension of the weights arrays
is equal to the length of longest input data. These arrays (also
referred as \emph{reference vectors}) determine the SOM behaviour.

To accomplish the classification, SOM goes through three phases:
initialization, training and classification.

\paragraph{Initialization} 
First of all, some system parameters (number of nodes, learning rate and
radius) have to be fixed by e.g. the IDS technician. The number of
nodes directly determines the classification given by the SOM: a small
network will classify different data inputs in the same node while a
large network will produce a too sparse classification. Afterwards,
the array of node weights is initialized, usually with random values
(in the same range of input values).

\paragraph{Training} 
The training phase consists of a number of iterations (also called
\emph{epochs}). At each iteration one input vector $x$ is processed as
follows: $x$ is compared to all neuron weight arrays $w_n$ with a
distance function (Euclidean or Manhattan): the most similar node
(also called \emph{best matching unit}, BMU) is then identified.

After the BMU has been found, the neighbouring neurons and the BMU
itself are updated. The following update parameters
are used: the neighbourhood is governed by the \emph{radius} parameter
($r$) and the magnitude of the attraction is affected by the
\emph{learning rate} ($\alpha$). 

During this phase, the map tends to converge to a stationary
distribution, which approximates the probability density function of
the high-dimensional input data.

As the learning proceeds and new input vectors are given to the map,
the learning rate and radius values gradually decrease to zero.

\paragraph{Classification} 
During the classification phase, the first part of the training phase
is repeated for each sample: the input data is compared to all the
weight arrays and the most similar neuron determines the
classification of the sample (but weights are not updated). The
winning neuron is then returned.

\subsection{PAYL classification model}

PAYL, is a n-gram~\cite{Dam95} analysis algorithm, and uses a
classification method based on clustering of packet payload data
length.

PAYL classifies packets on the \emph{length of the payload}. During
the training phase, for a given training data set, PAYL computes a set
of \emph{models} $M_{ijk}$. For each incoming packet, with destination
address $j$ and destination port $k$ and payload length $i$, $M_{ijk}$ stores incrementally the
average byte frequency and the standard deviation of each byte
frequency.  During the detection phase, the same values are computed
for incoming packets and then compared to model values: a significant
difference from the norm produces an alert. To compare models, PAYL
uses a simplified version of the Mahalanobis distance, which has the
advantage of taking into account not only the average value but also
its variance and the covariance of the variables measured.
 
The maximum amount of space required by PAYL is: $p * l * k$, where
$p$ is the total number of ports monitored (each host may have
different ports), $l$ is the length of the longest payload and $k$ is
a constant representing the space required to keep the mean and the
variance distribution values for each payload byte (PAYL uses a fixed
value of 512).

To reduce the otherwise large number of models to be computed, PAYL
organizes models in clusters. After comparing two neighbouring models
using the Manhattan distance, if the distance is smaller than a given
threshold $t$, models are merged: the means and variances are updated
to produce a new combined distribution. This process is repeated until
no more models can be merged. Experiments with PAYL show~\cite{WS04}
that a reduction in the number of model of up to a factor of 16 can be
achieved.

\paragraph{Modification to PAYL}

Our modification to PAYL works as follows: we pre-process each packet,
using the SOM. Afterwards PAYL uses the class value given by the SOM
(\emph{winning neuron}) instead of the payload length. Technically
PAYL, instead of using model $M_{ijk}$, uses the model $M_{njk}$ where $j$ and 
$k$ are the usual destination address and port and $n$ is the classification
derived from the neural network. Then, mean and variance values are
computed as usual.

Having added SOM to the system we must allow for both the SOM
and PAYL to be trained separately. Regarding resource consumption,
we have to revise the required amount of space to: $p * n * k$, where
the new parameter $n$ indicates the amount of SOM network nodes.

\section{Tuning and Experiments\label{sec:results}}
In this section, we show the results of our benchmarks and compare the
performance of POSEIDON with PAYL and PHAD. PAYL and PHAD are the two
reference ADS based on payload. They are the only two
ABS based on payload which have published their detection rate on the
DARPA 1999 data set.

\subsection{SOM parameters tuning}
The SOM algorithm needs several parameters on start-up: the total
number of network nodes, the function used to compute the distance
between vectors and the values of the \emph{learning rate} and
\emph{update radius}. For the sake of transparency, we report here the
values used in our experiments.

Concerning the number of neurons, a small network would yield a too
course classification, while a large network will produce a sparse
classification. In addition, it is worth bearing in mind that the
computational load increases quadratically with the number neurons.

Experimenting with different initialization parameters and using the
\emph{quantization error} method~\cite{Koh95} to evaluate the
classification given by the network, we found the best SOM with the
following parameters:

\begin{itemize}
\item Number of neurons: 96 (rectangular\\ network of 12 by 8).
\item Learning rate: 0.1.
\item Update radius: 4.
\item Distance function: Manhattan.
\end{itemize}

Hinneburg et al.~\cite{HAK00} state that Manhattan distance performs better than Euclidean 
distance in presence of high-dimensional data: our experiments substantially confirm this statement 
also in the case of network data analysis.

\begin{figure*}[ht]
\centering
  \includegraphics[height=12cm]{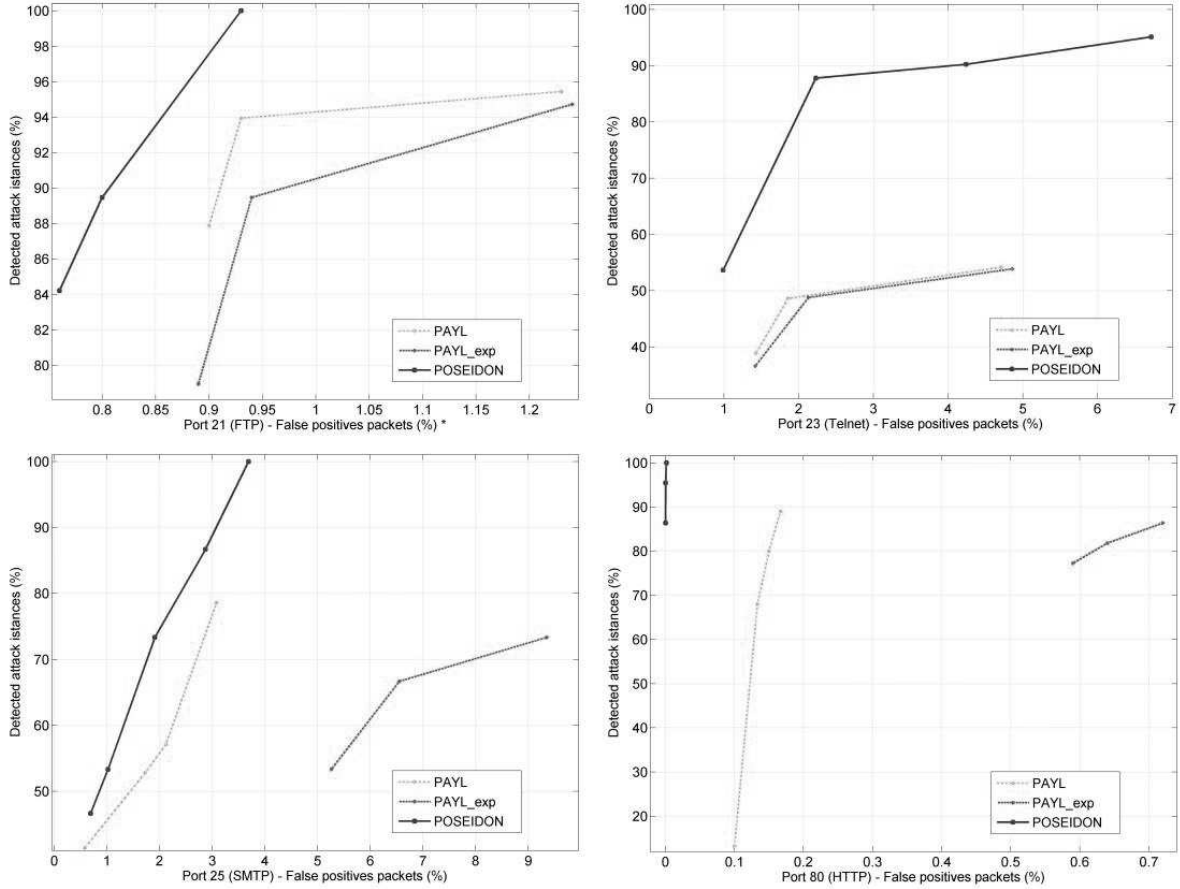}
\caption{Detection rates for ports 21 (FTP), 23 (Telnet), 25 (SMTP) and 80 (HTTP): the x-axis and y-axis present false positive rate and detection rate respectively. POSEIDON presents always a higher detection rate compared with PAYL at the same false positive rate. For the graph relative to port 21 see Remark in Section~\ref{par:ftp_remark}.\label{fig:graphs}}
\end{figure*}

\begin{table*}[ht]
\centering

\begin{tabular}{||r|c|c|c|c||}
\hline

\multicolumn{2}{||c|}{}& \textbf{PAYL} & \textbf{PAYL\_exp} &  \textbf{POSEIDON} \\
\hline

\multicolumn{2}{||c|}{Number of profiles used} & 4065 & (11312 - unclustered) & 1622 \\ 
\hline
\hline

\multirow{2}{*}{HTTP} & DR & 89,00\% & 90,00\% & 100,00\% \\
& FP & 0,17\% & 0,73\% & 0,0016\% \\
\hline

\multirow{2}{*}{FTP} & DR & 95,50\% & 94,74\% & 100,00\% \\
& FP & 1,23\% & 11,41\% (1,21\%$^*$) & 11,31\% (0,93\%$^*$) \\
\hline

\multirow{2}{*}{Telnet} & DR & 54,17\% & 53,65\% & 95,12\% \\
& FP & 4,71\% & 4,94\% & 6,72\% \\
\hline

\multirow{2}{*}{SMTP} & DR & 78,57\% & 73,34\% & 100,00\% \\
& FP & 3,08\% & 8,35\% & 3,69\% \\
\hline
\hline

\multicolumn{2}{||c|}{Overall DR with FP $<$ 1\%} & 58,8\% (57/97) &  & 73,2\% (71/97)$^*$ \\ 
\hline

\end{tabular}
\caption{Comparison between PAYL, our implementation of PAYL (PAYL\_exp) and POSEIDON; DR stands for detection rate, while FP is the false positive rate\label{tab:results1}}
\end{table*}


\begin{table*}[ht]
\centering

\begin{tabular}{||l|l|c|c||}
\hline

\textbf{Type} & \textbf{Attack} & \textbf{PHAD} & \textbf{POSEIDON} \\
\hline
Probe & ntinfoscan & 66,67\% (2/3) & 100\% (3/3) \\
\hline
\hline
\multirow{3}{*}
{Denial of Service} & apache2 & 100\% (3/3) & 100\% (3/3) \\
& back & 0\% (0/4)& 100\% (4/4) \\
& crashiis & 71,43\% (5/7) & 100\% (7/7) \\
\hline
\hline
\multirow{2}{*}{Remote to Local} & phf & 66,67\% (2/3) & 100\% (3/3) \\
& ppmacro & 33,34\% (1/3) & 100\% (3/3) \\
\hline
\hline
\multicolumn{2}{||l|}{Overall detection rate} & 65\% (13/20) & 100\% (20/20) \\
\hline
\end{tabular}
\caption{Comparison between PHAD and POSEIDON detection rates.\label{tab:results2}}
\end{table*}

\subsection{Experiments}
We have benchmarked POSEIDON against PAYL (also by replicating the
experiment on PAYL) and PHAD, using the same data used by PAYL and
PHAD: the DARPA 1999 data set \cite{LHFK+00}. This standard data set is
used as reference by a number of researchers
(e.g.~\cite{MC02,Ngu02,WS04}), and offers the
possibility of comparing the performance of various IDS. This data set
has been criticized because of the environment in which data were
collected~\cite{McH00}; as explained by Mahoney and Chan~\cite{MC03},
it is possible to tune an IDS in such a way that it scores
particularly well on this particular data set: some attributes --
specifically: remote client address, TTL, TCP options and TCP window
size -- have a small range in the DARPA simulation, but have a large
and growing range in real traffic. IDS which take into account the
above-mentioned attributes are likely to score much better on the
DARPA set than in real life. Since our system does not consider these
attributes, we can legitimately expect that the system in real life
performs as well as it does on the DARPA benchmark.

To compare our model with PAYL, we apply the same restrictions and
conditions used by Wang and Stolfo~\cite{WS04}: we focus only on inbound
TCP packets, with data payload, directed to hosts 172.016.0.0/16 and
ports 1-1024. 

We train the SOM clustering algorithm using internal
network traffic of week 1 and week 3 (12 days, 2.444.591 packets,
attack free): for each different protocol we use a different SOM.
Then, we use the same data to build PAYL models taking advantage of
the classification given by the neural network.

After this double training phase, it is possible to use the testing
weeks (4 and 5) to benchmark the network intrusion detection algorithm. This
data contains several attack instances (97 payload-based attacks are
detectable applying the same traffic filter mentioned above), as well
as legal traffic, directed against different hosts of the internal
network: the attack source can be situated both inside and outside the
network.

Figure~\ref{fig:graphs} shows a detailed comparison of PAYL and
POSEIDON in terms of percentage of true negatives (reported on the $y$
axis) w.r.t.~the percentage false positives ($x$ axis).
Table~\ref{tab:results1} reports a summary of these results: the first
column reports PAYL's statistics as we have inferred them from the
graphs reported by Wang and Stolfo~\cite{WS04}. The second column
reports the figures we obtained by repeating Wang and Stolfo's
benchmarks. In the repeated PAYL experiments we used an \textbf{un-clustered} architecture, which yields on one hand to a
higher number of profiles, and on the other hand to a different
classification. The third column reports POSEIDON's result.
Is it possible to observe that POSEIDON overcomes PAYL on every benchmarked protocol: there is a remark about FTP protocol (see the next paragraph).

\paragraph{Remark\label{par:ftp_remark}}During FTP protocol benchmarks we found a high rate of false positives
(more than 3000 packets) both with PAYL and with POSEIDON: all these
packets are sent by the same source host, which is sending FTP
commands in a way that is typical of the Telnet protocol (one
character per packet, with the TCP flag \emph{PUSH} set). These
packets are marked as an attack because the training model does not
contain this kind of traffic over the FTP control channel port,
although it is normal traffic. During our experiments with PAYL we
found the same behaviour: for this reason we decided to present
benchmarks results of PAYL and POSEIDON also without taking into account
these packets (the figures marked with an asterisk $^*$ in Table~\ref{tab:results1} and the graph in Figure~\ref{fig:graphs}).

Table~\ref{tab:results2} compares our results with PHAD: it is not
possible to make a full comparison between the two systems, because of
the restrictions used by PHAD authors (they restrict to a maximum
total amount of 100 false positives during 10 days of testing).
Nonetheless, we could legitimately compare the two systems on the HTTP
protocol, on which POSEIDON meets the restrictions above.

Unfortunately, there is no other public available data set suitable to
compare our approach with previous researches on anomaly intrusion
detection: many authors use the KDD 99 data set~\cite{BKPS00} in which
regrettably payload data is discarded. Because we use payload
information, we can not use this data set to benchmark POSEIDON and
models that use this data set are not directly comparable with ours.

Concluding, the significant achieved improvement over
PAYL is determined by a better distribution of mean and variance
value within categories, obtained with introduction of a new
classification algorithm (SOM).

\section{Related work}
\label{sec:related}
Network intrusion detection systems based on anomaly detection have been
widely studied for two decades. We recall that anomaly detection
systems can operate in various manners, sometimes extracting features
from packet headers and sometimes from payload data.

In this section we report on related work. First we describe other
neural network-based systems then we address statistical-based
systems.

\subsection{Neural networks based systems}

We start by presenting other neural-network based IDS. We cannot
benchmark these systems with POSEIDON because their authors use either
private data sets (Cannady~\cite{Can98}, Labib and Vemuri~\cite{LV02}
and Ramadas et al.~\cite{ROT03}), or data sets that do not contain
payload information (Depren et al.~\cite{DTAC04}) or do not provide
precise statistics (Nguyen~\cite{Ngu02}).

Cannady~\cite{Can98} proposes a SOM-based IDS in which network packets
are first classified according to nine features and then presented to
the neural network. Attack traffic is generated using a security audit
tool. The author extends this work in Cannady~\cite{Can00a,Can00b}.

Nguyen~\cite{Ngu02} uses a one-tier architecture, consisting of a SOM,
to detect two attacks in the 1999 DARPA data set: the first one
(\textit{mailbomb}) against the SMTP service, and the other one
(\textit{guessftp}) against FTP.

Labib and Vemuri~\cite{LV02} use a SOM to identify Denial of Service
attacks. They discard information about payload and use only packet
header information; their data is collected from a private network
(described in a general way) and is not publicly available.

Ramadas et al.~\cite{ROT03} use a SOM to detect attacks against DNS
and HTTP services (using a private data set): they use a pre-processor
to summarize some connection parameters (source and destination host
and port) and then add several values to track connections behaviour:
the information is then merged in a data structure used to fire events
related to the connection and to feed the neural network.

Depren et al.~\cite{DTAC04} present a hybrid IDS based on
self-organizing maps and benchmark it on the KDD 99 data
set~\cite{BKPS00}. They feed the neural networks (one for each
protocol type) with six features extracted from each connection
(duration, protocol type, service type, status, total bytes sent and
received) and then use the quantization error method to detect
anomalies. The system is connection-oriented, therefore attacks can be
detected only when the connection is completely re-assembled.
Regarding their architecture, the authors state that the SOM used to
model TCP connections uses 1515 neurons; which in our opinion is quite
large, if compared with the ones used by our system.

\subsection{Statistical-based systems}
In addition to ADS based on neural networks, there exist ADS employing
statistical models to detect anomalous behaviour.  We now report on
them. Again, we cannot benchmark them against POSEIDON because they
either use only header information (Hoagland~\cite{Hoa00}, Javitz and
Valdes~\cite{JV94a}) or employ benchmarking data that is not publicly
available (Kruegel et al.~\cite{KTK02}).

Barbar\'a et al.~\cite{BCJW01a, BCJW01b} use data mining techniques to detect attacks on network infrastructures: their system ADAM first applies association rules techniques to identify abnormal events in traffic data; then a classification algorithm is used to classify the abnormal events into normal instances and abnormal instances. The original work has been expanded in~\cite{BJW01c}.
Lee et al.~\cite{LS98, LSM99} propose a comprehensive framework based on data mining. For a complete overview of data mining techniques applied to intrusion detection see Julisch~\cite{Jul02}.

The SPADE~\cite{Hoa00}, NIDES~\cite{JV94a} and PHAD~\cite{MC02}
systems rely on statistical models computed on normal network traffic:
they work by extracting features from the packet header fields and
trigger an alarm when they recognize a significant deviation from the normal
model; most of the features extracted are related to IP addresses
(source and destination), destination service port and TCP connection
state (PHAD uses up to 34 attributes coming from Ethernet, IP and
application layer protocols packets). Our approach differs from the
one mentioned here in the following aspects: (a) it is payload-based:
we use only destination address and service port numbers to build a
profile for each port monitored, without taking care of other header
features (of the above systems only PHAD considers payload
information, we have compared it with our system in the previous
section). (b) We have a two-tier architecture in which the SOM is used
only to pre-process information.

Shifting to payload-based systems, Kruegel et al.~\cite{KTK02} show
that it is possible to find the description of a system that computes
a payload byte distribution and combines this information with
extracted packet header features: they first sort the resultant ASCII
characters by frequency and then aggregate them into six groups. As
argued by Wang and Stolfo~\cite{WS04}, this leads to a very course classification of
the payload.
 
PAYL works in a way similar to Kruegel et al.~\cite{KTK02} but models
the full byte distribution based on payload data length and operates a
clustering phase to cover possible missing lengths. The PAYL
architecture is made up of a single tier, while our architecture has
two different layers: the first one, made up by a SOM, is delegated to
classify packets only using payload data information, without using
payload length value.  The second layer is a modified version of PAYL
that computes byte distribution models using the classification
information coming from the first layer and extracting destination IP
address and service port from packets header.

Zanero~\cite{Zan05b} presents a two-tier payload-based system that combines a 
self-organizing map with a modified version of SmartSifter~\cite{YTWM00}. 
While this architecture is similar to POSEIDON, a full comparison is not 
possible because the benchmarks in \cite{Zan05b} concern only the FTP service 
an no details are given about experiments execution.
A two-tier architecture for intrusion detection is also outlined in 
Zanero and Savaresi~\cite{ZS04}.

\section{Conclusion}
\label{sec:conclusion}
We present an approach to Network Intrusion Detection that involves the
combination of two different techniques: a self-organizing map and the
PAYL architecture. We modify the original PAYL to take advantage of
the unsupervised classification given by the SOM, which then functions
as pre-processing stage.

Our experiments on the DARPA set show that our approach reduces the
number of profiles used by PAYL (payload length can vary between 0 and
1460 in a Local Area Network, while the SOM neural network used in our
experiments has less than one hundred nodes). Our experiments show that PAYL
without SOM requires 3 times as many profiles as with the SOM
pre-processing (see Table \ref{tab:results1}).

We benchmark POSEIDON extensively against the PAYL algorithm and data
sets showing a higher detection rate and lower false positives rate.

\paragraph{Acknowledgments}
We thank Herbert Bos for his valuable comments.

\appendix

\section{Appendix: POSEIDON inner functions}
\label{sec:appendix}
In this section we describe the inner mathematical functions and
algorithms used by POSEIDON.

\subsection{SOM algorithm}
\begin{tabbing}

DA\=TA \= TY\=PE\=\\
\\
$\mathit{RR~=~[0.0..255.0]}$\\
\>/* Reals (Double) between 0.0 and 255.0 */\\
$\mathit{l~=~length~of~the~longest~packet~payload}$\\
$\mathit{PAYLOAD~=~array~[1..l]~of~[0..255]}$\\
\\
DATA STRUCTURE\\
\\
$\mathit{N~=~non-empty~finite~set~of~neurons}$\\
\\
$\mathit{for~each~n~\in~N~let}$\\
\>$\mathit{w_n~:=~array~[1..l]~of~RR}$\\
\> \> \>/* array of weights associated */\\
\> \> \>/*  to each neuron $n$ */
\\
$\mathit{\alpha_0~\in~\mathbb{R}}$ \> \> \>/* Initial learning rate */\\
$\mathit{\alpha~:=~\alpha_0}$ \> \> \>/* Current learning rate */\\
$\mathit{r_0~\in~\mathbb{R}}$ \> \> \>/* Initial radius */\\
$\mathit{r~:=~r_0}$ \> \> \>/* Current radius */\\
$\mathit{\tau~\in~\mathbb{N}}$ \> \> \>/* Number of training epochs */\\
$\mathit{k~\in~\mathbb{N}}$ \> \> \>/* Smoothing factor */\\
\\
INIT PHASE\\
\\
$\mathit{for~each~n~\in~N}$\\
\> $\mathit{for~i~:=~1~to~l}$\\
\> \> $\mathit{w_n[i]~:=~random(RR)}$\\
\> \> \>/* Initialize with values in $RR$ */\\
\\
TRAINING PHASE\\
\\
INPUT:\\
\> $\mathit{x_t: PAYLOAD}$\\
\\
$\mathit{for~t~:=~1~to~\tau}$\\[2mm]
\> /* Find winning neuron */\\
\> $\mathit{win\_dist~:=~+\infty}$\\
\> $\mathit{win\_neuron~:=~n_0}$\\[2mm]
\> $\mathit{for~each~n~\in~N~do}$\\
\> \>$\mathit{dist~:=~manhattan\_dist(x_t,w_n)}$\\
\> \> $\mathit{if~(dist~\leq~win\_dist)~then}$\\
\> \> \> $\mathit{win\_dist~:=~dist}$\\
\> \> \> $\mathit{win\_neuron~:=~n}$\\
\> \> $\mathit{end~if}$\\
\> $\mathit{done(for)}$\\[2mm]
\>/* Process neighbouring neurons */\\
\>$\mathit{N_n~=~\{n~\in~\mathbb{N}~|~trig\_dist(n,win\_neuron)~\leq~r\}}$\\[2mm]
\>$\mathit{for~each~n_n~\in~N_n}$\\
\> \>$\mathit{for~i~:=~1~to~l}$\\
\> \> \>$\mathit{w_{n_n}[i]~:=~w_{n_n}[i] + \alpha * (w_{n_n}[i] - x_{t}[i])}$\\[2mm]
\> $\mathit{\alpha := \alpha_0 * \frac{k}{k+t}}$\\
\> $\mathit{r := r_0 * \frac{\tau - t}{\tau}}$\\[2mm]
$\mathit{done(for)}$\\
\\
CLASSIFICATION PHASE\\
\\
INPUT:\\
\> $\mathit{x:~PAYLOAD}$\\
\\
OUTPUT:\\
\> $\mathit{win\_neuron~\in~\mathbb{N}}$\\
\\
$\mathit{win\_dist~:=~+\infty}$\\
$\mathit{dist~:=~win\_dist}$\\
$\mathit{win\_neuron~:=~n_0}$\\[2mm]
$\mathit{for~each~n~\in~N~do}$\\
\> $\mathit{dist~:=~manhattan\_dist(x,w_n)}$\\
\> $\mathit{if~(dist~\leq~win\_dist)~then}$\\
\> \> $\mathit{win\_dist~:=~dist}$\\
\> \> $\mathit{win\_neuron~:=~n}$\\
\> $\mathit{end~if}$\\
$\mathit{done(for)}$\\[2mm]
$\mathit{return~win\_neuron}$\\
\end{tabbing}


\subsection{PAYL algorithm}
\begin{tabbing}
DA\=TA \= TY\=PE\\
\\
$\mathit{feature~vector~=~RECORD~[}$\\
\> \>$\mathit{mean~=~array~[1..256]~of~Real,}$\\
\> \> \>/* average byte frequency */\\
\> \>$\mathit{stdDev~=~array~[1..256]~of~Real}$\\
\> \> \>/* standard deviation of each */\\
\> \> \>/* byte frequency */\\
\>$\mathit{]}$\\[2mm]
$\mathit{profile~=~RECORD~[}$\\
\> \>$\mathit{ip~\in~\mathbb{N},}$ /* destination host address */\\
\> \>$\mathit{sp~\in~\mathbb{N},}$ /* destination service port */\\
\> \>$\mathit{fv~=~finite~set~of~n~feature~vectors}$\\
\>$\mathit{]}$\\
\> /* for each port monitored a profile */\\
\> /* with n feature vectors is associated */\\
\\
DATA STRUCTURE\\
\\
$\mathit{P~=~set~of~finite~profiles}$\\
$\mathit{threshold~\in~\mathbb{R}}$\\
\> /* numeric value used for anomaly */\\
\> /* detection given by user */\\
\\
TRAINING PHASE\\
\\
INPUT:\\
\> $\mathit{ip:~IP~address~\in~\mathbb{N}}$\\
\> $\mathit{sp:~service~port~\in~\mathbb{N}}$\\
\> $\mathit{n:~SOM~classification}$\\
\> $\mathit{x:~PAYLOAD}$\\
\\
$\mathit{for~each~p~\in~P~do}$\\
\>$\mathit{if~(p.ip~=~ip~and~p.sp~=~sp)~then}$\\
\> \>$\mathit{fv~=~p.getFV(n)}$\\ 
\> \>/* get feature vector with index n */\\
\> \>$\mathit{fv.update(x)}$\\
\> \>/* update byte frequency distributions */\\
\>$\mathit{end~if}$\\
$\mathit{done(for)}$\\
\\
TESTING PHASE\\
\\
INPUT:\\
\> $\mathit{ip:~IP~address~\in~\mathbb{N}}$\\
\> $\mathit{sp:~service~port~\in~\mathbb{N}}$\\
\> $\mathit{n:~SOM~classification}$\\
\> $\mathit{x:~PAYLOAD}$\\
\\
OUTPUT:\\
\> $\mathit{isAnomalous:~BOOLEAN}$\\
\> \>/* is the packet anomalous ? */\\
\\
$\mathit{dist~:=~+\infty}$\\
$\mathit{isAnomalous~:=~FALSE}$\\
\\
$\mathit{for~each~p~\in~P~do}$\\
\>$\mathit{if~(p.ip~=~ip~and~p.sp~=~sp)~then}$\\
\> \>$\mathit{fv~:=~p.getFV(n)}$\\ 
\> \> \>/* get feature vector with index n */\\
\> \>$\mathit{dist~:=~fv.getDistance(x)}$\\
\> \> \>/* get the distance between input */\\
\> \> \>/* data and associated profile */\\
\>$\mathit{end~if}$\\
$\mathit{done(for)}$\\[2mm]
$\mathit{if~(dist~\geq~threshold)~then}$\\
\> $\mathit{isAnomalous~:=~TRUE}$\\
$\mathit{end~if}$\\[2mm]
$\mathit{return~isAnomalous}$\\
\end{tabbing}

\bibliographystyle{latex8}
\bibliography{bibliography}

\begin{thebibliography}{10}\setlength{\itemsep}{-1ex}\small

\bibitem{And80}
J.~P. Anderson.
\newblock Computer {S}ecurity {T}hreat {M}onitoring and {S}urveillance.
\newblock Technical report, James P Anderson Co.\, Fort Washington, PA, April
  1980.

\bibitem{BCJW01b}
D.~Barbar\'a, J.~Couto, S.~Jajodia, and N.~Wu.
\newblock {ADAM}: a testbed for exploring the use of data mining in intrusion
  detection.
\newblock {\em SIGMOD Record}, 30(4):15--24, 2001.

\bibitem{BCJW01a}
D.~Barbar\'a, J.~Couto, S.~Jajodia, and N.~Wu.
\newblock {ADAM}: {D}etecting {I}ntrusions by {D}ata {M}ining.
\newblock In {\em IAW '01: Proc.~2nd IEEE SMC Information Assurance Workshop},
  2001.
\newblock URL http://www.itoc.usma.edu/Workshop/2001/Authors/
  Submitted\_Abstracts/paperT1A3(21).pdf.

\bibitem{BJW01c}
D.~Barbar\'a, J.~Couto, S.~Jajodia, and N.~Wu.
\newblock {D}etecting {N}ovel {N}etwork {I}ntrusions using {B}ayes
  {E}stimators.
\newblock In {\em SIAM '01: Proc.~1st SIAM International Conference on Data
  Mining}, 2001.
\newblock URL http://www.siam.org/meetings/sdm01/pdf/ sdm01\_29.pdf.

\bibitem{BKPS00}
S.~D. Bay, D.~Kibler, M.~Pazzani, and P.~Smyth.
\newblock The {UCI KDD} archive of large data sets for data mining research and
  experimentation.
\newblock {\em SIGKDD Exploration: Newsletter of SIGKDD and Data Mining},
  2(2):81--85, 2000.

\bibitem{Can98}
J.~D. Cannady.
\newblock Artificial neural networks for misuse detection.
\newblock In {\em NISSC' 98: Proc.~21st National Information Systems Security
  Conference}, pages 443--456, 1998.

\bibitem{Can00a}
J.~D. Cannady.
\newblock {\em An adaptive neural network approach to intrusion detection and
  response}.
\newblock PhD thesis, Nova Southeastern University, 2000.

\bibitem{Can00b}
J.~D. Cannady.
\newblock Next {G}eneration {I}ntrusion {D}etection: {A}utonomous
  {R}einforcement {L}earning of {N}etwork {A}ttacks.
\newblock In {\em NISSC '00: Proc.~23rd National Information Systems Security
  Conference}, 2000.
\newblock URL http://csrc.nist.gov/nissc/2000/proceedings/papers/033.pdf.

\bibitem{CCCR+05}
M.~Costa, J.~Crowcroft, M.~Castro, A.~Rowstron, L.~Zhou, L.~Zhang, and
  P.~Barham.
\newblock Vigilante: end-to-end containment of {I}nternet worms.
\newblock In {\em SOSP '05: Proc.~20th ACM Symposium on Operating Systems
  Principles}, pages 133--147. ACM Press, 2005.

\bibitem{Dam95}
M.~Damashek.
\newblock Gauging similarity with n-grams: {L}anguage-independent
  categorization of text.
\newblock {\em Science}, 267(5199):843--848, 1995.

\bibitem{Den87}
D.~E. Denning.
\newblock An {I}ntrusion-{D}etection {M}odel.
\newblock {\em IEEE Transactions on Software Engineering}, SE-13(2):222--232,
  February 1987.

\bibitem{DTAC04}
M.~O. Depren, M.~Topallar, E.~Anarim, and K.~Ciliz.
\newblock Network {B}ased {A}nomaly {I}ntrusion {D}etection using {S}elf
  {O}rganizing {M}aps ({SOM}s).
\newblock In {\em SIU '04: Proc.~12th IEEE National Conference on Signal
  Processing and Applications}, pages 76--79, 2004.

\bibitem{HAK00}
A.~Hinneburg, C.~C. Aggarwal, and D.~A. Keim.
\newblock What {I}s the {N}earest {N}eighbor in {H}igh {D}imensional {S}paces?
\newblock In A.~E. Abbadi, M.~L. Brodie, S.~Chakravarthy, U.~Dayal, N.~Kamel,
  G.~Schlageter, and K.~Whang, editors, {\em VLDB '00: Proc.~26th International
  Conference on Very Large Data Bases}, pages 506--515. Morgan Kaufmann, 2000.

\bibitem{Hoa00}
J.~Hoagland.
\newblock {S}tealthy {P}ortscan \& {I}ntrusion {C}orrelation {E}ngine
  ({SPADE}), 2000.
\newblock URL http://www.silicondefense.com/software/spice/.

\bibitem{isc}
{SANS} {I}nstitute -- {I}nternet {S}torm {C}enter web site.
\newblock URL http://isc.sans.org/index.php?on=toptrends.

\bibitem{JV94a}
H.~S. Javitz and A.~Valdes.
\newblock The {NIDES} {S}tatistical {C}omponent {D}escription and
  {J}ustification.
\newblock Technical Report A010, SRI, 1994.

\bibitem{Jul02}
K.~Julisch.
\newblock {D}ata {M}ining for {I}ntrusion {D}etection: {A} {C}ritical {R}eview.
\newblock Research Report RZ 3398, {IBM} Zurich Research Laboratory, 8803
  Ruschlikon, Switzerland, February 2002.

\bibitem{Koh95}
T.~Kohonen.
\newblock {\em Self-Organizing Maps}, volume~30 of {\em Springer Series in
  Information Sciences}.
\newblock Springer, 1995.
\newblock (Second Extended Edition 1997).

\bibitem{KTK02}
C.~Kruegel, T.~Toth, and E.~Kirda.
\newblock Service specific anomaly detection for network intrusion detection.
\newblock In {\em SAC '02: Proc.~2002 ACM Symposium on Applied Computing},
  pages 201--208. ACM Press, 2002.

\bibitem{LV02}
K.~Labib and V.~R. Vemuri.
\newblock {NSOM}: {A} {T}ool {T}o {D}etect {D}enial {O}f {S}ervice {A}ttacks
  {U}sing {S}elf-{O}rganizing {M}aps.
\newblock Technical report, University of California, Davis, 2002.

\bibitem{LS98}
W.~Lee and S.~Stolfo.
\newblock Data mining approaches for intrusion detection.
\newblock In {\em Proc.~7th {USENIX} Security Symposium}, pages 79–--94. USENIX
  Association, 1998.

\bibitem{LSM99}
W.~Lee, S.~J. Stolfo, and K.~W. Mok.
\newblock A {D}ata {M}ining {F}ramework for {B}uilding {I}ntrusion {D}etection
  {M}odels.
\newblock In {\em S\&P '99: Proc.~20th {IEEE} Symposium on Security and
  Privacy}, pages 120--132, 1999.

\bibitem{LHFK+00}
R.~Lippmann, J.~W. Haines, D.~J. Fried, J.~Korba, and K.~Das.
\newblock The 1999 {DARPA} off-line intrusion detection evaluation.
\newblock {\em Computer Networks: The International Journal of Computer and
  Telecommunications Networking}, 34(4):579--595, 2000.

\bibitem{MC02}
M.~V. Mahoney and P.~K. Chan.
\newblock Learning nonstationary models of normal network traffic for detecting
  novel attacks.
\newblock In {\em KDD '02: Proc.~8th ACM SIGKDD International Conference on
  {K}nowledge {D}iscovery and {D}ata mining}, pages 376--385. ACM Press, 2002.

\bibitem{MC03}
M.~V. Mahoney and P.~K. Chan.
\newblock An {A}nalysis of the 1999 {DARPA}/{L}incoln {L}aboratory {E}valuation
  {D}ata for {N}etwork {A}nomaly {D}etection.
\newblock In G.~Vigna, C.~Kruegel, and E.~Jonsson, editors, {\em RAID '03:
  Proc.~6th Symposium on Recent Advances in Intrusion Detection}, volume 2820
  of {\em LNCS}, pages 220--237. Springer-Verlag, 2003.

\bibitem{McH00}
J.~McHugh.
\newblock {T}esting {I}ntrusion {D}etection {S}ystems: a critique of the 1998
  and 1999 {DARPA} intrusion detection system evaluations as performed by
  {L}incoln {L}aboratory.
\newblock {\em ACM Transactions on Information and System Security (TISSEC)},
  3(4):262--294, 2000.

\bibitem{Ngu02}
B.~V. Nguyen.
\newblock Self organizing map ({SOM}) for {A}nomaly {D}etection.
\newblock Technical report, Ohio University, 2002.

\bibitem{ROT03}
M.~Ramadas, S.~Ostermann, and B.~C. Tjaden.
\newblock Detecting {A}nomalous {N}etwork {T}raffic with {S}elf-{O}rganizing
  {M}aps.
\newblock In G.~Vigna, C.~Kruegel, and E.~Jonsson, editors, {\em RAID '03:
  Proc.~6th Symposium on Recent Advances in Intrusion Detection}, volume 2820
  of {\em LNCS}, pages 36--54. Springer-Verlag, 2003.

\bibitem{Roe99}
M.~Roesch.
\newblock {S}nort - {L}ightweight {I}ntrusion {D}etection for {N}etworks.
\newblock In {\em LISA '99: Proc.~13th USENIX Conference on System
  Administration}, pages 229--238. USENIX Association, 1999.

\bibitem{snort}
Snort {N}etwork {I}ntrusion {D}etection {S}ystem web site.
\newblock URL http://www.snort.org.

\bibitem{WCS05}
K.~W. G.~C. S.~J. Stolfo.
\newblock Anomalous {P}ayload-based {W}orm {D}etection and {S}ignature
  {G}eneration.
\newblock In A.~Valdes and D.~Zamboni, editors, {\em RAID '05: Proc.~8th
  International Symposium on Recent Advances in Intrusion Detection}, volume
  3858 of {\em LNCS}, pages 227--246. Springer-Verlag, 2006.

\bibitem{WS04}
K.~Wang and S.~J. Stolfo.
\newblock Anomalous {P}ayload-{B}ased {N}etwork {I}ntrusion {D}etection.
\newblock In E.~Jonsson, A.~Valdes, and M.~Almgren, editors, {\em RAID '04:
  Proc.~7th Symposium on Recent Advances in Intrusion Detection}, volume 3224
  of {\em LNCS}, pages 203--222. Springer-Verlag, 2004.

\bibitem{YTWM00}
K.~Yamanishi, J.~Takeuchi, G.~J. Williams, and P.~Milne.
\newblock On-line unsupervised outlier detection using finite mixtures with
  discounting learning algorithms.
\newblock In {\em KDD '00: Proc.~6th ACM SIGKDD international conference on
  Knowledge Discovery and Data Mining}, pages 320--324. ACM Press, 2000.

\bibitem{Zan05b}
S.~Zanero.
\newblock Analyzing {TCP} {T}raffic {P}atterns using {S}elf {O}rganizing
  {M}aps.
\newblock In F.~Roli and S.~Vitulano, editors, {\em ICIAP '05: Proc.~13th
  International Conference on Image Analysis and Processing}, volume 3617 of
  {\em LNCS}, pages 83--90. Springer-Verlag, 2005.

\bibitem{ZS04}
S.~Zanero and S.~M. Savaresi.
\newblock Unsupervised learning techniques for an intrusion detection system.
\newblock In {\em SAC '04: Proc.~19th Annual ACM Symposium on Applied
  Computing}, pages 412--419. ACM Press, 2004.

\end{thebibliography}

\end{document}